\documentclass[journal]{IEEEtran}
\usepackage{amsmath,amsfonts,amssymb}
\usepackage{algorithmic}
\usepackage{algorithm}
\usepackage{array}
\usepackage{textcomp}
\usepackage{stfloats}
\usepackage{url}
\usepackage{verbatim}
\usepackage{graphicx}
\usepackage{subcaption}
\usepackage{cite}
\begin{document}

\title{MAC Revivo: Artificial Intelligence Paves the Way}

\author{
        Jinzhe Pan,
        Jingqing Wang,~\IEEEmembership{Member,~IEEE,}
        Zelin Yun,
        Zhiyong Xiao,
        Yuehui Ouyang,
        Wenchi Cheng,~\IEEEmembership{Senior Member,~IEEE},
        and Wei Zhang,~\IEEEmembership{Fellow,~IEEE}
\thanks{Jinzhe Pan, Zelin Yun, Zhiyong Xiao and Yuehui Ouyang are with the Wireless Communication Department, Honor Device Co. Ltd., Shenzhen, China (e-mails: panjinzhe@honor.com; yunzelin@honor.com; xiaozhiyong@honor.com; yuehuiouyang@honor.com).}
\thanks{Jingqing Wang and Wenchi Cheng are with the State Key Laboratory of Integrated Services Networks, Xidian University, Xi’an, China (e-mails: jqwangxd@xidian.edu.cn; wccheng@xidian.edu.cn).}
\thanks{Wei Zhang is with the School of Electrical Engineering and Telecommunications, The University of New South Wales, Sydney, Australia (e-mail: w.zhang@unsw.edu.au).}
}


\maketitle

\begin{abstract}

The vast adoption of Wi-Fi and/or Bluetooth capabilities in Internet of Things (IoT) devices, along with the rapid growth of deployed smart devices, has caused significant interference and congestion in the industrial, scientific, and medical (ISM) bands. 
Traditional Wi-Fi Medium Access Control (MAC) design faces significant challenges in managing increasingly complex wireless environments while ensuring network Quality of Service (QoS) performance. 
This paper explores the potential integration of advanced Artificial Intelligence (AI) methods into the design of Wi-Fi MAC protocols. We propose AI-MAC, an innovative approach that employs machine learning algorithms to dynamically adapt to changing network conditions, optimize channel access, mitigate interference, and ensure deterministic latency.
By intelligently predicting and managing interference, AI-MAC aims to provide a robust solution for next generation of Wi-Fi networks, enabling seamless connectivity and enhanced QoS. 
Our experimental results demonstrate that AI-MAC significantly reduces both interference and latency, paving the way for more reliable and efficient wireless communications in the increasingly crowded ISM band.
\end{abstract}


\section{Introduction}
\IEEEPARstart{W}{ith} the rapid advancement of wireless communications technologies, next-generation Wi-Fi networks promise significant performance enhancements and new features, aiming for ultra-low latency ($<$1 ms), high reliability ($>99.99999\%$), and improved energy efficiency~\cite{10177877}. However, the increasing demand for these metrics, along with the proliferation of intelligent Internet of Things (IoT) devices, introduces challenges in network performance and design. 
Traditional Medium Access Control (MAC) protocols struggle to meet these advanced requirements in congested and interference-prone wireless environments, particularly in the industrial, scientific, and medical (ISM) bands. 

Conventional Wi-Fi networks typically optimize parameters in isolation, and do not have an integrated, end-to-end approach. 
To address these limitations, there is growing interest in integrating Artificial Intelligence (AI) into MAC protocol design, known as AI-MAC. 
AI-MAC leverages advanced machine learning (ML) algorithms to improve throughput, optimize network utilization, and ensure Quality of Service (QoS) and Quality of Experience (QoE) in dynamic and heterogeneous environments~\cite{ABBASI2021104234,6913496}. 
By adapting to real-time conditions, optimizing channel access, and mitigating interference, AI-MAC has the potential to achieve deterministic latency and enhance network performance, although implementation at large scale remains challenging.

Recently, extensive research has explored the application of learning models to enhance Wi-Fi network performance, highlighting the pivotal role of AI in improving MAC functionalities. 
Recent surveys highlight the need for structured approaches to analyze performance metrics of Wi-Fi networks using ML techniques~\cite{9786784}. 
Key areas of focus include performance optimization, channel selection mechanisms, network management, indoor localization, and quality metrics related to QoS/QoE~\cite{9855457,GUERIN2023129}.
AI models are instrumental in extracting valuable features from large-scale network data streams. 
For instance, the authors of~\cite{li2024} introduced an ML-based scheduling framework that optimizes application-layer QoS in Wireless Local Area Networks (WLANs) experiencing co-channel interference. Furthermore, Multi-Agent Reinforcement Learning (MARL) algorithms have been proposed to replace inefficient random channel access mechanisms with deep reinforcement learning-based protocols~\cite{9681886,10339653}, enabling deterministic decision-making and enhanced QoS delivery.

While these advances highlight the potential of AI-MAC, the progresses remains largely incremental, and focused on isolated aspects within the traditional paradigm. 
The evolution of AI in Wi-Fi is still in its early stages, accumulating knowledge and waiting for a significant breakthrough~\cite{10634004}. 
In summary, the road to commercialization remains long and challenging. 
In contrast, AI has achieved remarkable success in fields such as Computer Vision (CV) and Natural Language Processing (NLP), revolutionizing applications and gaining widespread acceptance. 
This disparity raises the question: why is AI progress lagging behind in wireless communication? 
Despite the obvious need for AI integration in communications systems, the industry has yet to find an effective approach.

\begin{figure*}[htbp]
  \vspace{-11pt}\centering
  \includegraphics[width=5.3in]{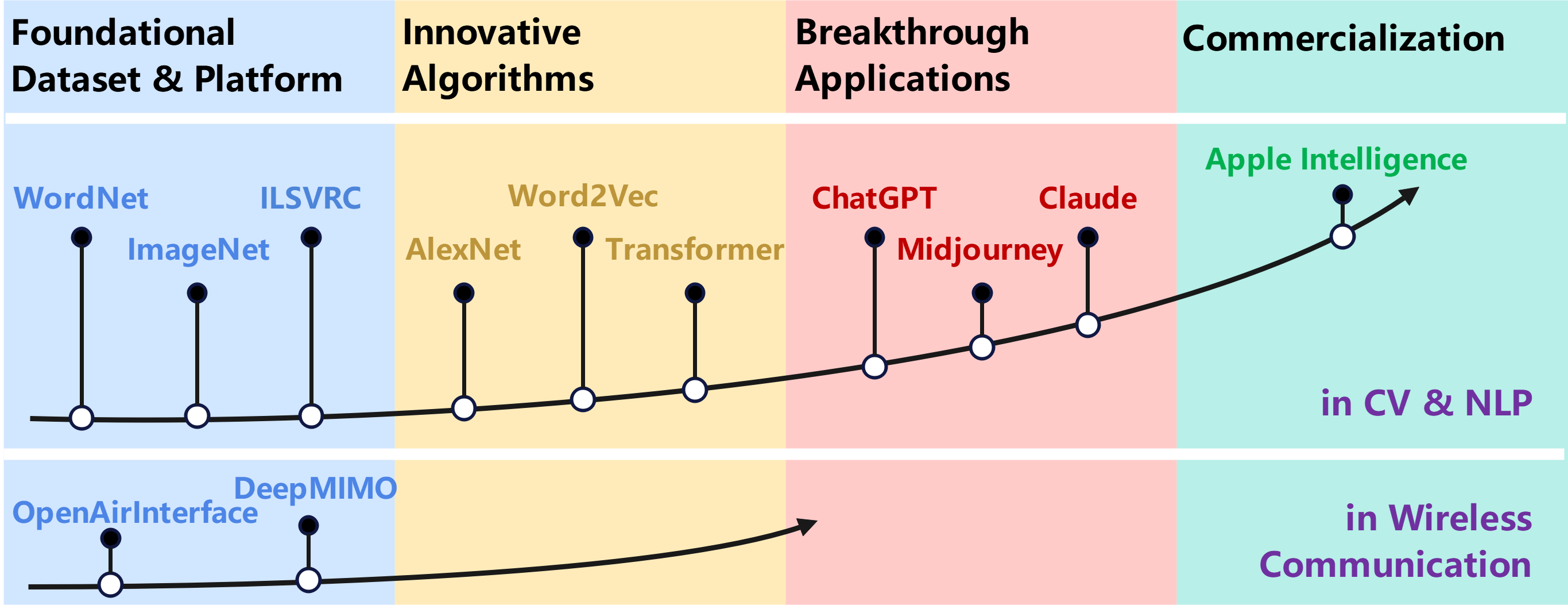}
  \caption{Advances of AI in CV, NLP, and wireless communication}
  \label{fig_1}
  \vspace{-9pt}
\end{figure*}

We draw parallels to the rapid advances of AI in CV and NLP, and propose a roadmap, shown in Fig.~\ref{fig_1}, to expedite AI-driven progress:
\begin{itemize}
  \item{\bf Foundational Datasets and Platform:} The initial step, inspired by the way ImageNet and its Challenges (ILSVRC) revolutionized the CV field, is to establish public datasets and a platform for algorithm validation. 
  Current datasets in wireless communications focus largely on cellular networks and Wi-Fi localization, leaving a gap in MAC protocol datasets. 
  Given the hardware dependency in wireless systems, making direct performance comparisons difficult, a simulation platform with standardized datasets will spur further research.
  \item{\bf Innovative Algorithms:} The accumulation of research work eventually leads to groundbreaking algorithms like AlexNet.
  As The Economist noted, “Suddenly people started to pay attention, not just within the AI community but across the technology industry as a whole.” 
  This paradigm shift is inspiring a wave of new research and applications, driving substantial progress in the field.
  \item{\bf Breakthrough Applications:} Major algorithmic innovations, such as \textit{word2vec} and \textit{Transformer}, have led to historic breakthroughs. The recent success of \textit{ChatGPT} demonstrates the commercial potential of such advances, and is creating a global sensation. 
  This has led to a flood of various applications. Everyone is eagerly anticipating a similar breakthrough in the communications field. 
  \item{\bf Commercialization:} The culmination of these advances will attract a broad research community and promote collaboration between academia and industry. 
  By building a robust ecosystem for AI-driven innovation, we can accelerate the integration of AI into wireless communication networks and realize their full potential.
\end{itemize}

This paper aims to explore the transformative potential of AI-MAC for next-generation Wi-Fi networks. 
We propose an AI-MAC framework that integrates AI into MAC protocol design and discuss the fundamental principles and advantages of AI-MAC over traditional MAC protocols. 
More importantly, we develop a research platform and model library built on ns-3, designed to replicate interference environments using real-world data. 
Our goal is to unify currently fragmented research efforts by providing a standardized, open platform for researchers. 
We call for more algorithms to foster community growth and accelerate progress in the field, ultimately bringing the `ChatGPT moment' closer in the AI-MAC domain.

\section{AI-MAC Framework}
\begin{figure*}[htbp]
    \vspace{-9pt}\centering
  \includegraphics[width=6.7in]{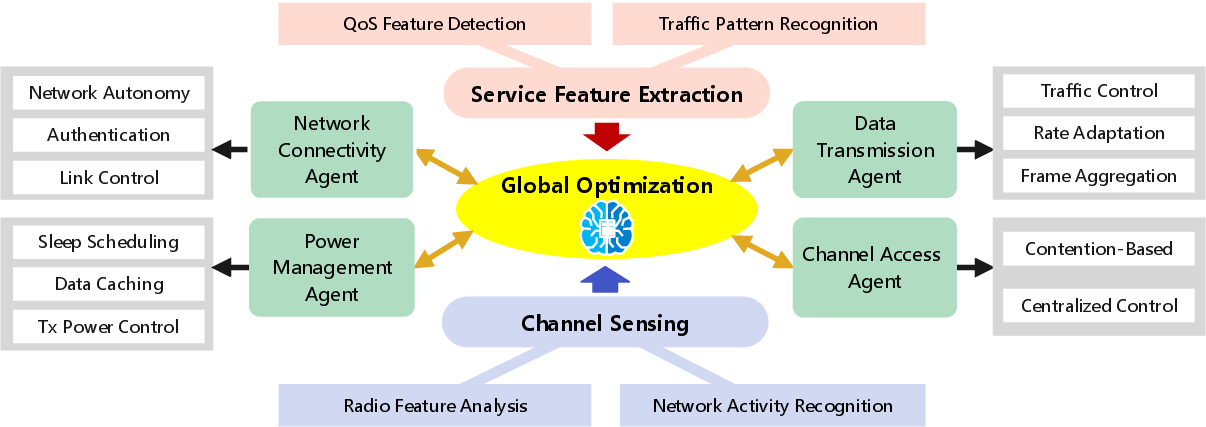}
  \caption{AI-MAC Framework}
  \label{fig_2}  \vspace{-9pt}
\end{figure*}

In this section, we introduce the core features of the AI-MAC framework, as shown in Fig. \ref{fig_2}, and provide an overview of the key design challenges and potential solutions for integrating AI models in Wi-Fi environments.

\subsection{Enhancing Conventional Wi-Fi Functionalities}
We examine how AI/ML can enhance four key functionalities in Wi-Fi networks.

\subsubsection{Channel Access}
Distributed Coordination Function (DCF) mechanisms like CSMA/CA are crucial for Wi-Fi but become inefficient in dynamic, dense environments due to static access strategies, leading to increased collisions and delays. 
AI can improve DCF by enabling adaptive decision-making, where ML algorithms dynamically adjust parameters like backoff times and contention windows. 
Reinforcement learning (RL) can optimize these parameters by learning from real-time network conditions, reducing collisions and enhancing throughput.
Similarly, the Hybrid Coordination Function (HCF) 
aims to enable contention-free access but face challenges with complexity and scalability. AI can optimize these systems through dynamic scheduling, where ML models predict traffic patterns and allocate time slots efficiently. 

\subsubsection{Data Transmission}
The AI-MAC framework enhances data transmission through intelligent traffic control mechanisms.
It dynamically prioritizes network traffic based on data type and QoS requirements, ensuring high-priority traffic receives sufficient bandwidth and experiences minimal delay. 
The framework also includes an AI-driven queue scheduler that monitors and reorganizes packet queues, ensuring reliable user experiences. 
Rate adaptation strategies involve dynamic modulation and coding scheme (MCS) selection and interference mitigation, continuously optimizing transmission parameters for throughput, delay, and reliability. 
AI models further improve bandwidth efficiency through frame aggregation, selecting optimal frame sizes to meet diverse QoS needs.

\subsubsection{Network Connectivity}
Network connectivity management encompasses key functions such as network autonomy, authentication, and link control. 
The complexity arises from the interdependence of links, where adjustments to one link affect both its performance and that of neighboring devices, especially in dense environments. 
By analyzing real-time data and historical patterns, the framework can forecast network behavior and prepare appropriate adjustments in advance. 
This approach reduces the likelihood of outages and ensures continuous optimization of network configurations.

\subsubsection{Power Management}
By calibrating power levels in response to real-time network conditions and QoS/QoE requirements, the framework ensures that each device utilizes only the necessary amount of power for reliable connectivity, thereby conserving energy and reducing interference. Additionally, AI-MAC optimizes the utilization of wake-up modes in network devices. AI-driven models analyze traffic patterns to predict low-activity periods, intelligently transitioning devices into sleep states during these intervals. This strategy minimizes unnecessary power consumption without compromising network's responsiveness and overall performance.

\subsection{Innovative Modules for Enhanced Network Intelligence}
The AI-MAC framework introduces two innovative modules aimed at enhancing network intelligence and performance by providing comprehensive data for AI decision-making. 

\subsubsection{Service Feature Extraction}
Service feature extraction involves identifying and analyzing key characteristics of various services to optimize management and resource allocation strategies. 
QoS feature detection gathers real-time data to extract metrics such as latency, jitter, and reliability~\cite{10609803}, enabling better prioritization of network resources.
Traffic pattern recognition categorizes traffic types like streaming media, file transfers, and online gaming based on data flow characteristics. 
By recognizing these patterns, the framework predicts future network conditions and proactively manage resources to prevent congestion and ensure efficient data transmission. 
This information facilitates adaptive traffic management policies that uphold diverse QoS/QoE levels, allocating essential resources to high-priority, time-sensitive applications.

\subsubsection{Channel Sensing}
Channel sensing is a crucial component of AI-MAC framework, providing critical insights into the network environment. 
It encompasses two key functionalities: radio feature analysis and network activity recognition.
The radio feature analysis collects and analyzes various characteristics of the air interface, including signal power, traffic patterns, and channel utilization. 
This information enables the system to choose proper transmission power or switch to less congested channels, thereby improving resouce usage and reducing interference.
On the other hand, network activity recognition continuously monitors and recognizes the behaviors of devices within the network, which allows the system to infer the services and needs of these devices.
This capability not only helps forecast future conditions but also facilitates traffic prioritization and optimization of transmission parameters. As a result, critical applications receive the necessary resources, leading to an improved user experience. 

\subsection{Global Optimization}
A key advantage of the AI-MAC framework is its capability for global optimization, distinguishing it from traditional local optimization methods. 
The process begins with the collection and analysis of real-time data from active services and air interface, including factors such as traffic patterns, QoS requirements, and device behaviors. 
By integrating inputs from innovative modules previously discussed, the framework achieves a comprehensive understanding of the global state. This holistic perspective serves as a foundation for informed decision-making across the functional modules.

This global information acquired enhances the AI-MAC framework's capabilities in macro-level functions like network management. 
By synthesizing insights from various modules, the system can make informed decisions over a broader time scale, ensuring the stability and overall performance of the network.
Moreover, the framework also considers the distinct needs of different functional modules. 
Unlike current AI/ML methods that either focus solely on local optimization or attempt to optimize all functionalities as a single entity, the framework strikes a balance by allowing each module to operate independently while remaining interconnected.
After undergoing global training as a cohesive unit, each module is empowered to make its own decisions. 
This flexibility enables rapid decision-making for functionalities that require immediate responses, such as channel access. 
Overall, this holistic approach not only optimizes resource consumption but also minimizes response delays, ultimately leading to a more efficient and responsive network environment.



\section{Research Platform}
An open research platform with standardized datasets is crucial for accelerating technological developments in academic and industrial communities.  
In this section, we propose an AI-MAC research platform comprising following models. 
\begin{figure*}[htbp]
   \vspace{-9pt} \centering
  \includegraphics[width=6.6in]{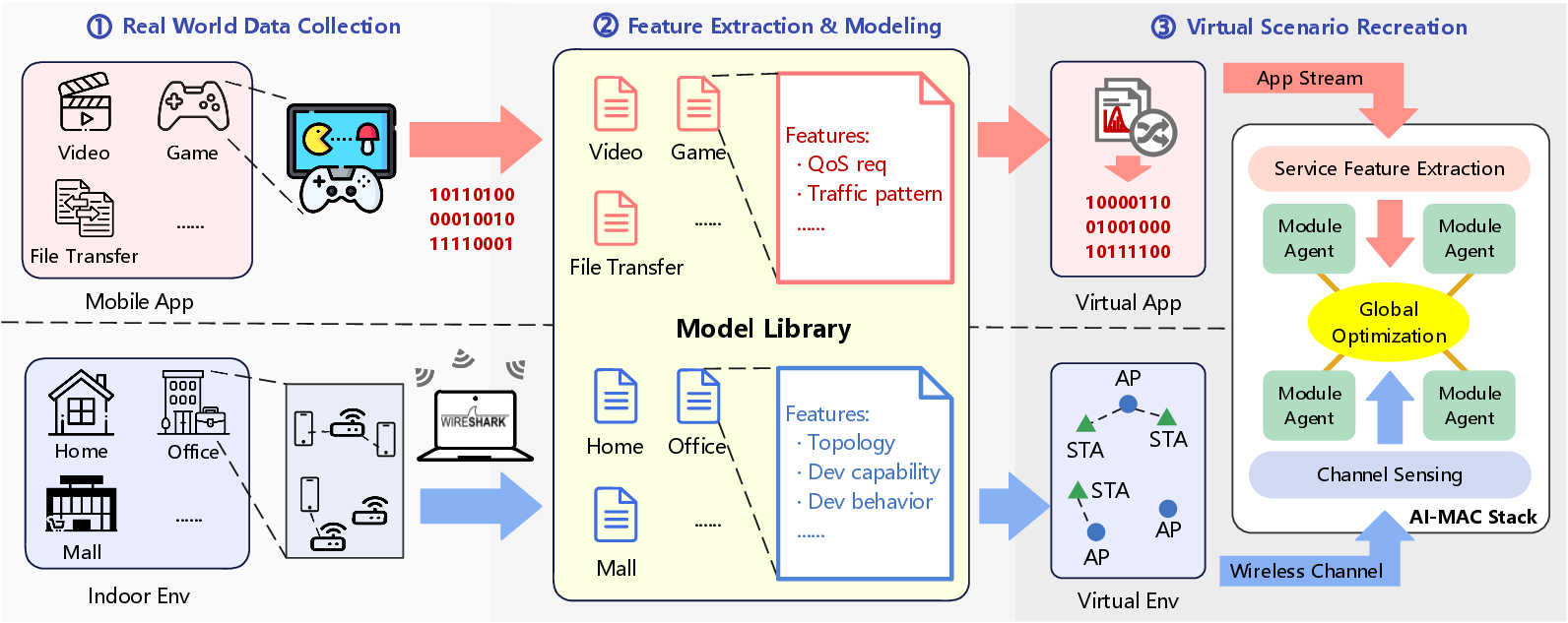}
  \caption{The construction and workflow of the proposed simulation platform}
   \vspace{-9pt} \label{fig_3}
\end{figure*}

\subsection{Interference Environment Model}


Building a comprehensive interference dataset is crucial for understanding and developing new technologies to mitigate interference in wireless communications.
Simply collecting real data and replaying it on a simulation platform is inadequate, as this approach fails to capture the interactive nature of interference.
Moreover, training AI models requires substantial amounts of data. 
The training and validation process can be significantly enhanced by using synthetic data, which can effectively reflect a variety of scenarios.

To address these challenges, we propose an interference environment model representing statistical behavior of real-world interfering devices.
As shown in Fig. \ref{fig_3}, we use Wireshark to capture Wi-Fi packets on 80MHz channels in 5GHz band at various representative locations, including homes, offices, shopping malls, airports, and train stations. The collected data is subsequently analyzed to extract information such as the number of APs and stations (STAs), their locations, association relationships, configurations, capabilities, and behaviors. 
We employ statistical models to fit the distribution of traffic from interfering devices and develop corresponding profiles, which are then used to recreate synthetic interference environments. 
This methodology enables us to generate a sufficient amount of synthetic data for training, resulting in a more dynamic and realistic simulation of interference scenarios.

\subsection{Application Traffic Model}
To effectively manage and optimize network resources, it is essential to understand the diverse traffic patterns and develop an application traffic model to support traffic generation in the simulation platform.
To build accurate application traffic models, we collect traffic data from a range of applications, including file transfer, video streaming, online gaming, and VoIP. 
This data is then analyzed to identify key characteristics of the traffic patterns, such as packet size and inter-arrival time, while also documenting their corresponding QoS requirements.
Similar to the interference environment model, we employ statistical models to fit the distribution of these characteristics. 
Based on these attributes of traffic profiles, we install virtual applications on the device that generate specific synthetic traffic flow in the simulator.

The interference environment model and the application traffic model together form a comprehensive model library. 
This library is used to generate configuration files for simulating various scenarios of interest, ensuring a realistic and dynamic evaluation environment.

\subsection{Simulation Platform}
The proposed simulation platform is built on \textit{ns-3} and \textit{ns3-ai} \cite{10.1145/3389400.3389404}. 
In this setup, the ns-3 simulator handles the network simulation tasks, while ns3-ai facilitates inter-process communication between C++ and Python, enabling seamless integration of AI frameworks such as \textit{PyTorch}. 

The AI-MAC protocol stack is established within the platform. 
It provides methods for acquiring information required by two innovative modules, and interfaces that facilitate efficient information exchange between functional modules and AI algorithms. 
Based on the profiles provided by the model library, the platform is capable of simulating a variety of user scenarios by deploying devices, installing virtual applications, and generating synthetic traffic data. 
In addition, the platform includes data collection and visualization tools that present performance metrics, channel occupancy, and collision data. This functionality enhances researchers' understanding of the underlying mechanisms of the system.
This setup enables researchers to focus solely on AI algorithm innovation. By providing a realistic and controlled environment, the platform allows for efficient development and testing of algorithms, thereby facilitating the advancement of AI-MAC solutions.

\section{Proposed AI-MAC Algorithm}
In this section, we introduce AI-MAC mechanism, adaptively managing and optimizing protocol stack parameters. 
This mechanism aims to provide agile, stable, and customized services with low-tail QoS requirements for delay/jitter/reliability-sensitive transmissions. 
Our preliminary algorithm demonstrates the potential of machine learning in optimizing MAC layer decisions.

\subsection{Mechanism of the Proposed Algorithm}
\begin{figure}[tbp]
  \vspace{-9pt}  \centering
  \includegraphics[width=3.1in]{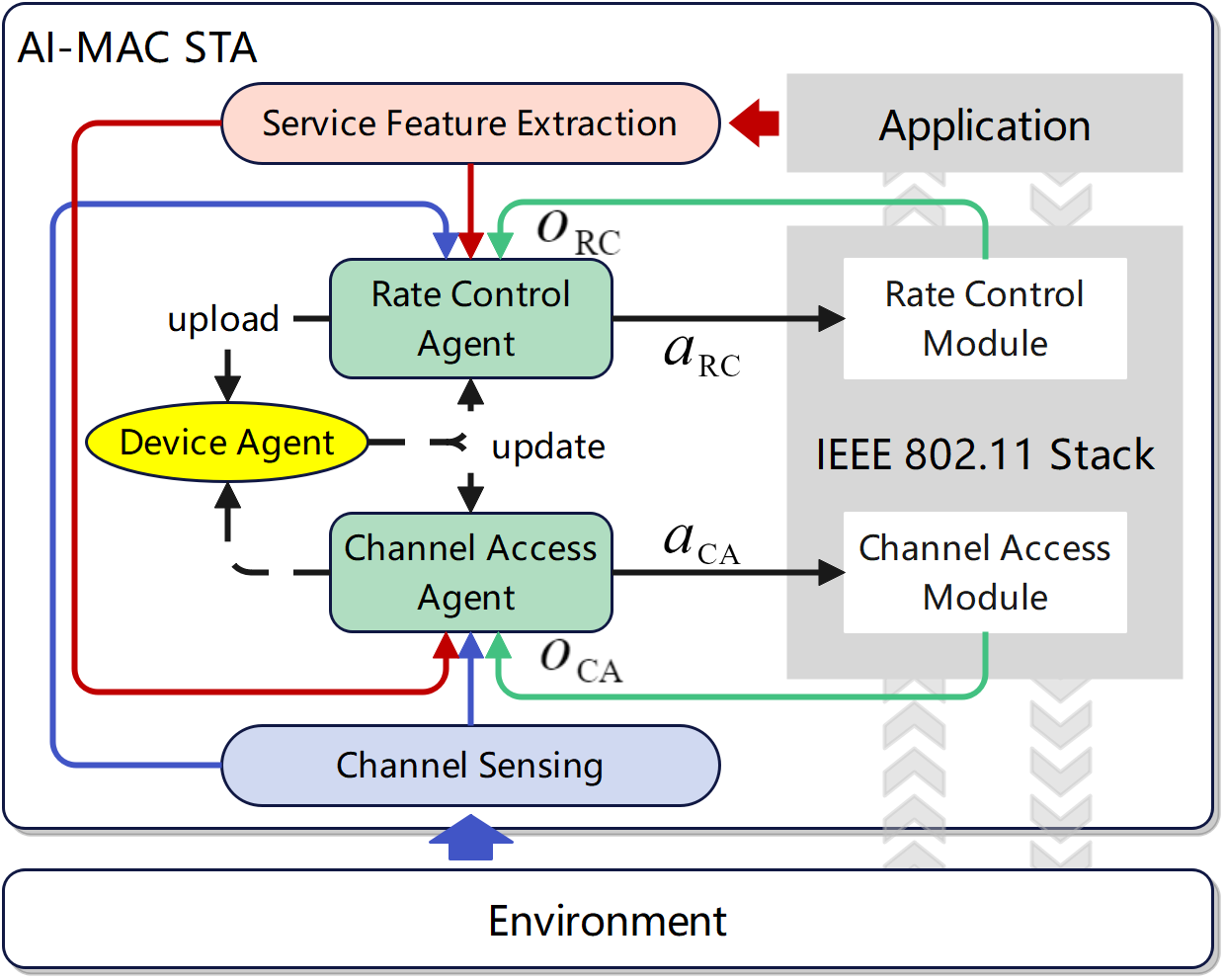}
  \caption{The framework of the proposed AI-MAC algorithm}
  \label{fig_4}  \vspace{-9pt}
\end{figure}

AI-MAC equips each functional module with an event-triggered module agent, as illustrated in Fig. \ref{fig_4}. 
The number and types of agents can be customized to meet user requirements, ensuring adaptability across diverse user scenarios. 
These agents operate asynchronously, which introduces challenges related to effective coordination among them while pursuing global optimization. 
To address these complexities, we employ an asynchronous MARL framework in conjunction with the Centralized Training with Decentralized Execution (CTDE) paradigm.
As shown in Fig. \ref{fig_4}, a device agent is introduced to coordinate among module agents, playing a central role in processing uploaded local information, organizing training, and facilitating model updates.
This agent constructs a comprehensive view of the global state, ensuring that the system effectively meets overall QoS requirements.

The channel access agent is triggered at the edge of each accessible idle time slot (i.e., the edge of the backoff countdown in traditional CSMA/CA).
While the rate control agent is activated during the generation of the TXVector for the data frame to be sent. When triggered, each module agent uploads its local observation, action, and reward to the device agent. 
QMIX \cite{JMLR:v21:20-081} as an effective network architecture under the CTDE paradigm is employed.
Since the module agents operate asynchronously, a single experience is constructed at the device agent based on each agent's most recent upload. 

\subsection{Local Observations and Actions}
As illustrated in Fig. \ref{fig_4}, each module agent independently collects local observations through its corresponding functional module, as well as from the channel sensing and service feature extraction modules. 
These observations can be divided into two parts: state information and module information.  
The state information includes the current trigger time and system details that reflect the global optimization objectives, while the module information pertains to specific functionalities.

For the channel access agent, it outputs an action $a_{\text{CA}}$, which is either \textit{Transmit} or \textit{Wait}.
Its observation $o_{\text{CA}}$ includes the channel busy indicator, the number of observed devices, and the delay to last successful transmission for current and interfering devices~\cite{9681886}. 
Moreover, it incorporates the \textit{waited slot counter}, a novel metric that increments with each \textit{Wait} decision and resets upon executing \textit{Transmit}. 
The rate control agent produces an output action $a_{\text{RC}}$ selected from the HE-MCS set.
Its observation $o_{RC}$ includes the recent SNR and the ratio of successfully transmitted frames in last sent packet.

\subsection{Global State and Reward}
The QoS requirements in the proposed AI-MAC are encapsulated in three key service features: delay-bounded, jitter-bounded, and error-rate bounded. 
These features utilize decay rates to measure the probability of QoS violations, with faster decay rates indicating a more effective system in reducing the likelihood of service degradation.
To effectively reflect the performance of these identified QoS features, real-time data -- such as delay, jitter and packet loss rate -- is collected from both the air interface and the device itself.
This data characterizes the current network condition and application requirements, forming an integral part of the state information. 
Together with the actions of the module agents, this state information defines the global state of the system.

Based on this global state, the state reward $r_{\text{state}}$ is designed to prioritize QoS for low-tail cases by evaluating violation probabilities.
This approach ensures efficient support for specific requirements rather than solely assessing average performance metrics~\cite{9324753}.
The state reward is expressed as:
\begin{equation}
	r_{\text{state}}=f(g(app,qosReq),h(intf)),
  \label{eq:r_state}
\end{equation}
where $g(\cdot)$ is the feature extraction module that processes application-specific parameters $app$ and the QoS features $qosReq$, $h(\cdot)$ represents the channel sensing module that collects and analyzes channel observation $intf$, and $f(\cdot)$ serves as the overall mapping function that compares actual performance against service requirements to calculate state reward. 

The local reward for each module agent is calculated based on the module information.
The channel access agent aims to utilize the channel efficiently while minimizing collisions and ensuring fairness among devices.
Its local reward $r_{\text{CA}}$ is designed as the sum of the transmission reward and the fairness reward.
This reward structure incentivizes successful transmissions and collision avoidance, while penalizing collisions resulting from ACK timeouts and prolonged idle wait times.
The fairness reward encourages equitable access to the channel among multiple devices.
The rate control agent's local reward $r_{\text{RC}}$ is based on the successful transmission ratio, ensuring the robustness of the transmission.

The global reward is then computed, taking into account the state reward in (\ref{eq:r_state}) and contributions from each individual module agent. 
To accommodate the asynchronous nature of the system, a discount factor $\gamma$ is applied, leading to the following expression for the total reward:
\begin{equation}
  r_{\text{tot}} = r_{\text{state}} + \sum_{i\in\{\text{CA}, \text{RC}\}} \gamma^{t-t_i} r_i,
  \label{eq:r_tot}
\end{equation} 
where $t$ is the current time and $t_i$ is the latest trigger time of the module agent $i$.

\section{Experimental Results}
In this section, we present superior simulation results of AI-MAC compared to 802.11ax and 802.11be, highlighting its effectiveness in real-time mobile gaming applications that require deterministic latency. 

\begin{figure*}[htbp]
   \vspace{-8pt} \centering
  \includegraphics[width=6.6in]{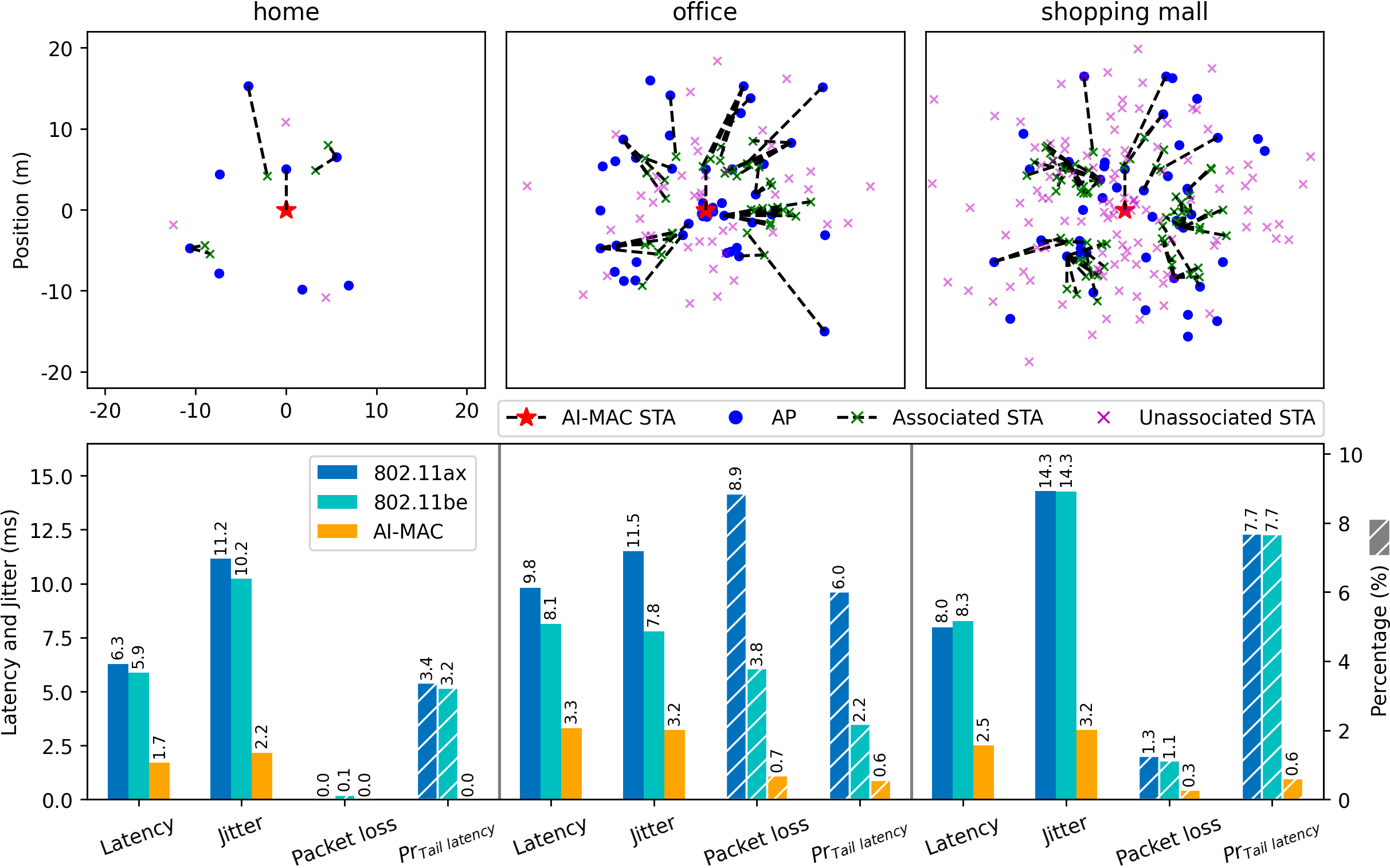}
  \caption{Performance comparison of Wi-Fi standards and AI-MAC under different scenarios}
  \label{fig_5}  \vspace{-8pt}
\end{figure*}

\subsection{Simulation Setup}
The real-time mobile gaming application is inherently delay-sensitive, and is attracting significant attention from game developers, device manufacturers, and network operators. 
In this work, we focus on a specific application characterized by traffic patterns outlined in the RTA report \cite{IEEE80211RTA}. 
Our primary interest lies in the traffic generated during gameplay, particularly concerning the status synchronization mechanism.

A STA equipped with the proposed AI-MAC algorithm is connected to an AP. 
The packet sizes and arrival times for both uplink and downlink traffic follow the largest extreme distributions.
While the throughput requirement is low -- 30-150 bytes every 10-30 milliseconds on average -- each packet carries critical user instructions and calculation results that must be delivered promptly.
Although uplink and downlink traffic are closely related in practice, we treat them to be independent of each other in this work.

To evaluate the performance of our algorithm, we replicate three scenarios: home, office, and shopping mall.
The device topologies in these scenarios are illustrated in Fig. \ref{fig_5}, where the device positions are randomly generated based on their RSSI values in the configuration file.
In the home scenario, the smaller area minimizes interference from neighboring devices, resulting in minimal interference. 
In contrast, the office scenario features a higher density of APs and associated STAs, leading to significantly increased throughput due to a large volume of aggregated data frames.
The shopping mall scenario presents a different dynamic: numerous APs are spread over a larger area but there are fewer associated STAs. Instead, there are a large number of unassociated STAs sending probing requests due to heavy foot traffic. As a result, the air interface is predominantly occupied by management and control frames, which leads to a lower proportion of data frames compared to the office scenario.

The module agents firstly undergo centralized training under the coordination of the device agent. 
During the performance evaluation, they make independent inference. 
To assess performance under varying conditions, we carry out 20 independent tests, each lasting 15 seconds, using different random seeds to simulate different interference environments and traffic flows.
The performance metrics are then averaged across all tests.

\subsection{Performance Metrics and Results}
The stability and reliability are critical, given the unique characteristics and QoS requirements associated with the mobile gaming application.
To evaluate the performance of the proposed algorithm, the following metrics are used:
\begin{itemize}
  \item{\bf Latency:} Latency represents the average delay of all successfully transmitted packets.
  \item{\bf Jitter:} Jitter is defined as the standard deviation of the delays of all successfullly transmitted packets.
  \item{\bf Packet loss:} Packet loss is the percentage of dropped packets relative to the total number of generated packets.
  \item{\bf Tail latency probability:} $Pr_{\text{Tail~latency}}$ is the percentage of packets with delays exceeding 30 milliseconds relative to the total number of successfully transmitted packets.
\end{itemize}

As illustrated in Fig. 5, 802.11ax and 802.11be struggle with reliability and stability,especially under strong interference. 
In contrast, AI-MAC effectively reduces latency and packet loss rate, while maintaining low jitter levels and tail latency. 
This enhances responsiveness and fluidity in application interactions, and ensures a stable and seamless gaming experience.

These results suggest that the integration of fairness reward and state reward in (\ref{eq:r_state}) enables AI-MAC to optimize its performance by adaptively managing channel access while minimizing interference with other users. 
This strategy allows the packets to be sent more promptly, effectively reducing latency and jitter.
The short packet nature inherent in mobile gaming ensures that the device only occupies limited channel resources, leading to balanced resource allocation across the network.
This adaptability indicates that AI-MAC has effectively learned the characteristics and requirements of the application, ensuring efficient resource usage while maintaining robustness.

\section{Future Research Directions}
In this paper, we propose a novel AI-MAC framework and detail its functionalities, research platform and initial algorithms. 
We have established a foundation for integrating AI into MAC protocols. 
Although our preliminary results are promising, they mark only the beginning of the journey toward full adoption of AI-MAC in future Wi-Fi networks. 
It is imperative to identify key research areas that address existing limitations and expand the capabilities of AI-MAC. 
The following discussion outlines several open challenges and suggests possible avenues for future research.

\subsection{Algorithmic Refinement and Scalability}
Future research will focus on enhancing the scalability and robustness of AI algorithms to manage the flexibility and complexity of next-generation Wi-Fi networks. 
This includes developing sophisticated AI models that is capable of handling wireless networks with multiple devices, adapting in real time to the dynamics of air interface, and balancing multiple QoS demands in complex user scenarios. 
More challenging high-throughput applications, such as AR/VR, will be studied to see how innovative AI-MAC algorithms can meet the demanding bandwidth and low-latency requirements simultaneously. 

\subsection{Dataset Expansion and Benchmarking}
The current model library requires continuous expansion and diversification to cover a wider range of scenarios, including those with higher user density, mobility, heterogeneous environments and different QoS requirements. 
We plan to expand data collection to include the 2.4GHz band and Bluetooth devices, enhancing the robustness and applicability of our models with more real-world data.
Additionally, benchmarking standards will be created to enable comparison of different AI algorithms to promote a collaborative research environment.

\subsection{Joint Cross-Layer Optimization of Diverse Wi-Fi Features}
It has become clear that to achieve optimal performance, diverse Wi-Fi features should be jointly optimized across multiple layers of the network. 
Future research will therefore focus on integrating AI-MAC protocols with other layers, such as the PHY and network layers, when managing various Wi-Fi functions. This joint cross-layer optimization approach is expected to lead to more holistic optimization strategies and enable seamless coordination between resource allocation, power control, delay-bound thresholds, routing decisions, etc.

\subsection{Real-World Implementation and Testing}
Developing testbeds that enable experimental research on AI-powered Wi-Fi networks is a crucial step before deploying AI-driven solutions in real-world environments. Therefore, future research will focus on going beyond simulations and collaborating with industry partners to validate the effectiveness of AI-MAC protocols under actual operating conditions. These efforts will also help identify practical challenges that may arise during deployment.




\bibliographystyle{IEEEtran}
\bibliography{refs}

\end{document}